\documentstyle[preprint,epsf,aps,prl]{revtex}
\tightenlines

\begin{document}
\draft
\preprint{}
\title{New Classes of Quasicrystals and Marginal Critical States}
\author{Nobuhisa Fujita and Komajiro Niizeki}
\address{
Department of Physics, Graduate School of Science,\\
Tohoku University, Sendai 980-8578, Japan
}
\date{\today}
\maketitle
\begin{abstract}
One-dimensional quasilattices, namely, the geometrical objects to represent
quasicrystals, are classified into mutual local-derivability (MLD) classes.
Besides the familiar class, there exist an infinite number of new
MLD classes, and different MLD classes are distinguished by the inflation
rules of their representatives. It has been found that electronic
properties of a new MLD class are characterized by the presence of marginal
critical states, which are considered to be nearly localized states.

\end{abstract}

Quasicrystals (QCs) have aperiodic ordered structures, which are different
from either of crystalline and disordered materials. It has been reported
that QCs exhibit quite unique physical properties \cite{St99}, which should
be closely connected with the structure of QCs. A sufficient understanding
in this respect is, however, yet to be attained. For instance, in order to
understand the transport properties of QCs, further systematic studies on
the electronic properties of various types of QCs are necessary.

The atomic positions of an idealized QC form a quasiperiodic object called
quasilattice (QL), the classification of which is the principal subject of
the crystallography of QCs. One of the important features of QLs is the
self-similarity, which is closely connected with the physical properties.
Thus, in the first part of this letter, we shall develop a new
classification scheme of QLs based on the self-similarity. One can argue
that the different classes of QLs correspond to different universality
classes with respect to the electronic properties. We next show that there
exists a new class of QLs whose electronic properties are
characterized by the presence of {\it marginal critical states}, which can
be thought of as being virtually localized. This feature is essentially
different from the case of the standard class of QLs on which most of the
previous studies have been done. The transport properties of the new class
may be markedly different from the conventional one.

We confine ourselves to one-dimensional (1D) cases. A 1D QL is obtained by
projecting a subset of a 2D lattice $\Lambda$, the mother lattice, onto a
1D subspace $E_{\parallel}$ \cite{DuKa85,Lu93}. The line $E_{\parallel}$
is taken to be parallel to $\tau{\bf a}_1+{\bf a}_2$ with ${\bf a}_1$ and
${\bf a}_2$ being the primitive lattice vectors of $\Lambda$ and $\tau$ a
quadratic irrational such as $\frac{1}{2}(1 + \sqrt{5})$ (the golden mean),
$1 + \sqrt{2}$ (the silver mean), etc. The subset is taken as $\Lambda \cap
\Sigma$ with $\Sigma$ being a parallel strip to $E_\parallel$. If the
scales of $E_\parallel$ and $E_{\perp}$, the physical space and its
orthogonal complement, are chosen appropriately, both the projections
$\Lambda_\parallel$ and $\Lambda_\perp$ of $\Lambda$ onto $E_\parallel$ and
$E_\perp$, respectively, are given by the dense set ${\it{\bf Z}}[\tau]
\equiv \{p + q\tau \;|\; p, q \in {\it{\bf Z}}\}$. Thus, a 1D QL is a
discrete subset of $\Lambda_\parallel$,
where its point density is proportional to the window $W$, i.e., the width
of the strip $\Sigma$. According to the relative position of $\Sigma$ to
$\Lambda$, an infinite number of QLs are possible with the same $W$. But
they form a single {\it local-isomorphism class}, and their differences are
irrelevant to the physical properties.

For a generic $W$, one obtains a ternary 1D QL composed of three types of
lattice spacings, s, m and l, referring to short, medium and long spacings,
respectively; they satisfy $|$l$|=|$s$|+|$m$|$. If $W$ is reduced, one
obtains another QL as a subset of the original QL. The new spacings, s$'$,
m$'$ and l$'$, are composites of the older ones. For an appropriate choice
of the reduction of $W$, the older QL turns out to be a uniform decoration
of the newer, i.e., the decoration of l$'$, for example, is common for all
l$'$s. The three decorations of the spacings form a substitution rule. The
newer QL is then called a {\it subquasilattice} (SQL) of the older.

If $W$ is reduced by the factor $\tau^{-n}$ with $n$ being any positive
integer, the resulting QL is locally isomorphic with the scaled version of
the original QL with the factor $\tau^n$ \cite{Ni89}. The original QL is
called self-similar if the new QL is an SQL of the original. Here, the
substitution rule, combining the three spacings of the original QL with
the newer ones, represents the self-similarity; it is called the {\it
inflation rule}. It can be shown that a QL can be self-similar if and only if
its window $W$ belongs to the quadratic field ${\it{\bf Q}}[\tau] \equiv
\{r + s\tau \;|\; r, s \in {\it{\bf Q}}\}$ \cite{KNNF}; that is, the window
must be rational. The inflation rule as well as the minimum power $n$ for the
ratio of self-similarity, $\tau^n$, depends on a number theoretical property
of $W$. The following argument will be focused on this rational case.

Several QLs associated with the silver mean are presented in Table
\ref{table1}. The first two are binary but others ternary. The QLs except
for ``C'' and ``E'' have self-similarities whose ratios are presented in
the 6th column. The QL ``C'' is not self-similar but is changed to the
once-inflated version of ``A'' by the substitution rule in the 7th column.
A similar relation holds between the pair $\{$D, E$\}$. The QL ``A'' is an
SQL of ``B'' with the substitution rule, s$'$ = s, l$'$ = sm, and the QL
``F'' is an SQL of ``G'' with s$'$ = m, m$'$ = l, l$'$ = msm. Though the
two QLs ``G'' and ``H'' are similar, there exists no inflation rule which
combine them.


We introduce an important relationship between QLs: two QLs are {\it
mutually locally-derivable} if all the sites of one of them are determined
{\it locally} from the structure of the other and vice versa \cite{Ba91}.
All the QLs derived from a single mother lattice are classified into an
infinite number of {\it mutual local-derivability} (MLD) classes. The MLD
class to which a given QL belongs is determined by a number theoretical
property of $W \in {\it{\bf Q}}[\tau]$. We only present a necessary
condition: {\it if two QLs belong to a single MLD class, the relevant two
windows have a common denominator when they are represented as simple
fractions in the quadratic field}. A QL and its any SQL belong to the same
MLD class. The eight QLs in Table \ref{table1} are divided into three MLD
classes: $\{$A, B, C$\}$, $\{$D, E$\}$, $\{$F, G, H$\}$. It can be shown
that i) every MLD class includes at least one self-similar member, which
can be taken as a representative of the class, and ii) every QL in the MLD
class has an SQL which is similar to the representative; that is, any
non-self-similar QL is a uniform decoration of a self-similar QL.

The structure factor of a QL is composed of Bragg peaks, whose intensities
are determined by the size of the window. For example, the three QLs, ``C'',
``D'' and ``F'' in Table \ref{table1}, have windows of similar sizes, and
their structure factors are not so much different. They nevertheless belong
to different MLD classes.

We shall call a QL to be type I or II according as its window belongs to
${\it{\bf Z}}[\tau]$ or not, respectively. All the type I QLs for a given
$\tau$ form a single MLD class, as seen in Table \ref{table1}. Previous
investigations on the electronic properties of 1D QLs have been almost
exclusively done on the basis of models on type I QLs
\cite{KST87,NiNo86,AsSt88}. It has been proved that a type I QL
and its any decoration belong to a common universality class of electronic
properties \cite{Ko92}, which are dominated by the structures of the
inflation rule. On the other hand, we may expect that type II QLs belong to
different universality classes, because their inflation rules are different
from that of the type I class.

We now present a brief review on the electronic properties of type I QLs.
The main observations, obtained from the case of the Fibonacci lattice as
well as some of its associates, are \cite{KST87}:
 i)  {\it the energy spectrum is purely singular-continuous or,
equivalently, fractal-like}, and
 ii) {\it all the eigenfunctions are critical, i.e., neither extended nor
localized in the usual meaning}.
The energy spectrum and the eigenfunctions have also been found to exhibit
self-similar structures, which can be directly related to the inflation
rule of the underlying QL by a real-space renormalization-group approach
\cite{NiNo86}.

The energy spectrum of any homogeneous 1D system obeys locally a scaling
law. The local scaling at the reference energy $E_{\rm r}$ is represented
by the scaling exponent $\alpha=\alpha(E_{\rm r})$ satisfying $0 \le \alpha
\le 1$ \cite{KST87}. While $\alpha$ represents the local dimension of
the energy spectrum, it also characterizes the localization character of
the eigenfunction of the energy level $E_{\rm r}$. In particular, an
isolated energy level of a localized state has a vanishing exponent,
$\alpha=0$, while inside an absolutely continuous spectrum, which confirms
extended states, $\alpha=1$. A purely singular-continuous spectrum being
characterized by fractional exponents confirms critical eigenfunctions. The
energy spectrum of a type I QL is, in general, a multifractal
\cite{HJKP86Ko88}, and is
characterized by the $f(\alpha)$ spectrum. The support of the $f(\alpha)$
is an interval $[\alpha_{min}, \alpha_{max}]$ with $0 < \alpha_{min} <
\alpha_{max} < 1$; $\alpha$ ranges from $\alpha_{min}$ to $\alpha_{max}$,
so that every energy level is characterized by fractional power-law scaling
\cite{KST87}.

\begin{figure}
\begin{center}\epsfile{file=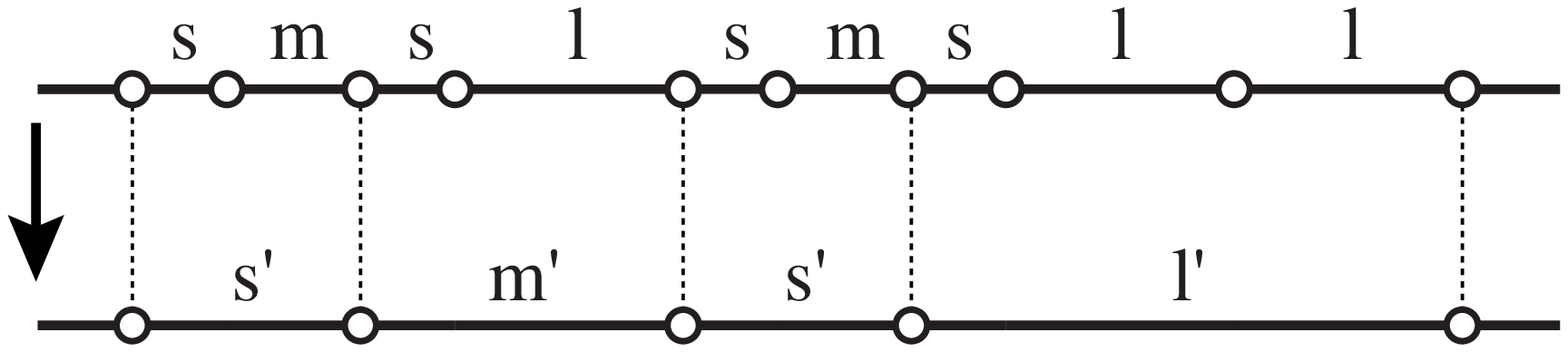,height=1.9cm}\end{center}
\caption{A ternary 1D QL given as ``D'' in Table I. It is self-similar with
the inflation rule in the table. The ratio of the self-similarity is the
silver mean, $\tau = 1 + \protect\sqrt{2}$. The frequencies of the three
spacings s, m, and l are shown to be proportional to
$\tau^{-1}:\tau^{-2}:\tau^{-1}$ ($2\tau^{-1} + \tau^{-2} = 1$).}
\label{fig:ql}\end{figure}

Although the above observations are common in the case of type I QLs,
electronic properties of type II QLs have been scarcely investigated. In
the following, we show some numerical as well as analytical evidence that
the type II QL listed as ``D'' in Table \ref{table1} and shown in Fig.\
\ref{fig:ql} indeed exhibits a new scaling property. Let us take a {\it
binary} atomic chain, which is obtained by decorating this QL as follows:
i) atoms of type X are located on all the lattice points, ii) one
atom of type Y is located on each spacing of type m, and iii) a pair of
type Y atoms is located on each spacing of type l. The densities of the two
types of atoms in the chain are equal. More remarkably, the chain is
invariant against the exchange of two types of atoms, X
and Y. We employ the tight-binding model on this atomic chain,
$t\Psi_{j-1}+V_j\Psi_j+t\Psi_{j+1}=E\Psi_j$, where we assume that $t=-1$
for the transfer integrals and $V_X=0$ and $V_Y=V$ for the relevant site
energies.

\begin{figure}
\begin{center}\epsfile{file=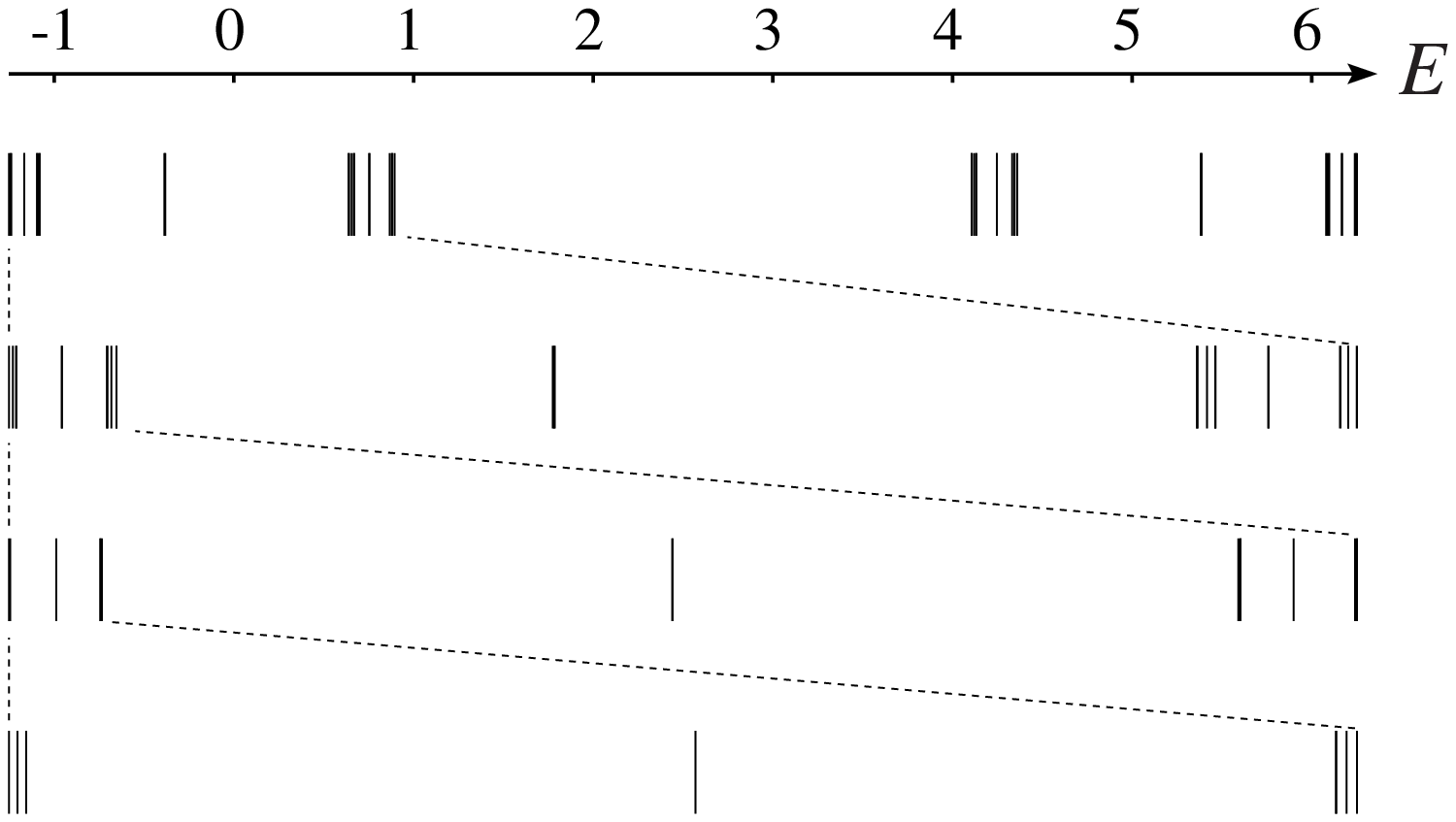,height=4.7cm}\end{center}
\caption{A singular-continuous energy spectrum of the type II QL given
in Fig.\ 1. It was obtained numerically from a finite approximant
composed of 1970 atoms with $V = 5.0$. The left cluster of energy
levels exhibits trifurcating behavior when it is expanded successively.}
\label{fig:spc}
\end{figure}
\begin{figure}
\begin{center}\epsfile{file=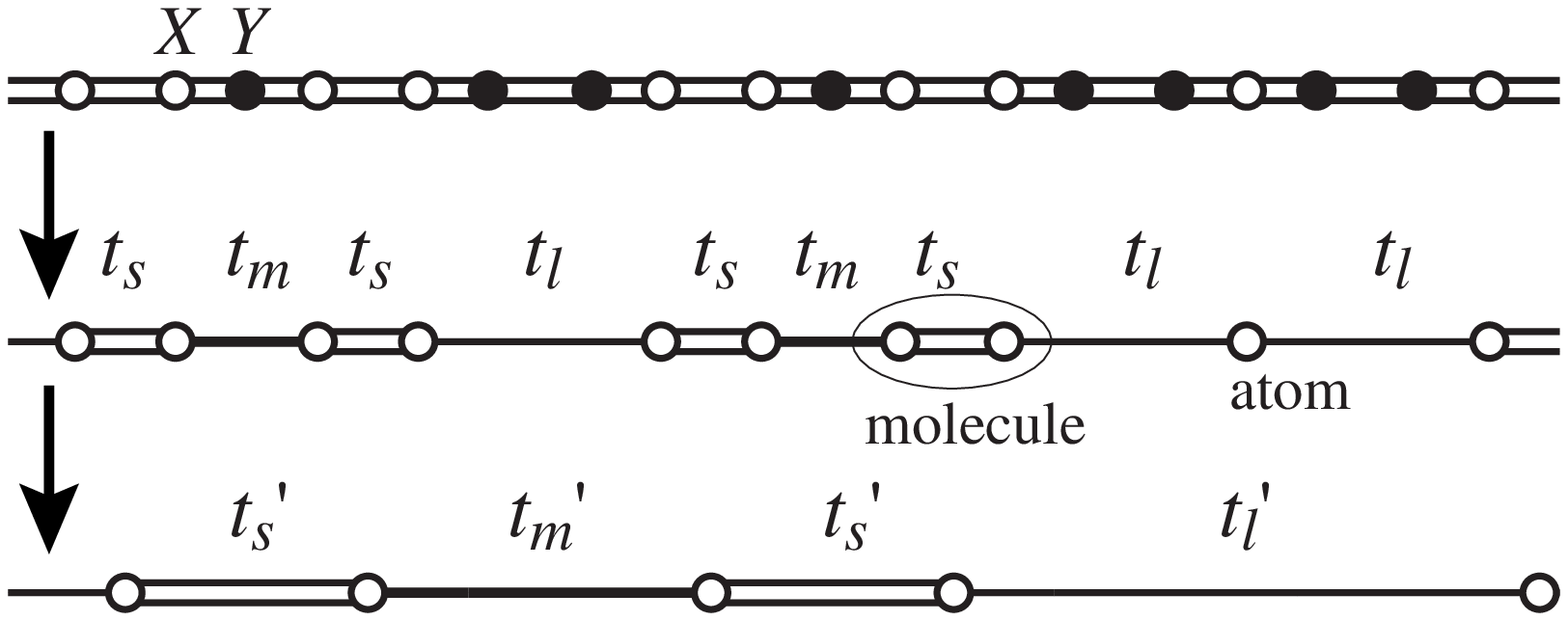,height=3.3cm}\end{center}
\caption{A perturbational real-space renormalization-group treatment of the
ground state. ``Molecules'' are separated by three types of spacings, which
are arranged similarly to the underlying QL.}\label{fig:dec}
\end{figure}

Due to the aforementioned symmetry of the atomic chain, the energy spectrum
becomes symmetrical which is confirmed by the numerical result as shown
in Fig.\ \ref{fig:spc}. The energy spectrum is divided into two equivalent
clusters, and the left cluster exhibits a hierarchical trifurcation, which
will reflect the nature of the leftmost level, i.e., the ground state. The
trifurcation can be understood in terms of a perturbational real-space
renormalization-group approach similar to that introduced by Niu and Nori
for the case of the Fibonacci lattice \cite{NiNo86}. The approach is based
on the recursive structure of the QL and assumes $V\gg 1$. In the zeroth
approximation, the spectrum consists of two equally degenerate energy
levels at $E=0$ and $V$, while the eigenstates are the Wannier states
themselves. To see the splitting of the level originating from X atoms, all
the Y atoms are decimated as shown in Fig.\ \ref{fig:dec}. This yields
three types of effective transfer integrals between X atoms,
$t_{\rm s}=-1$, $t_{\rm m} \approx -1/V$, and $t_{\rm l} \approx -1/V^2$,
which are all negative and satisfy the inequalities:
\begin{equation}
|t_{\rm s}| \gg |t_{\rm m}| \gg |t_{\rm l}|.
\label{inequality}\end{equation}

\begin{figure}[t]
\begin{center}\epsfile{file=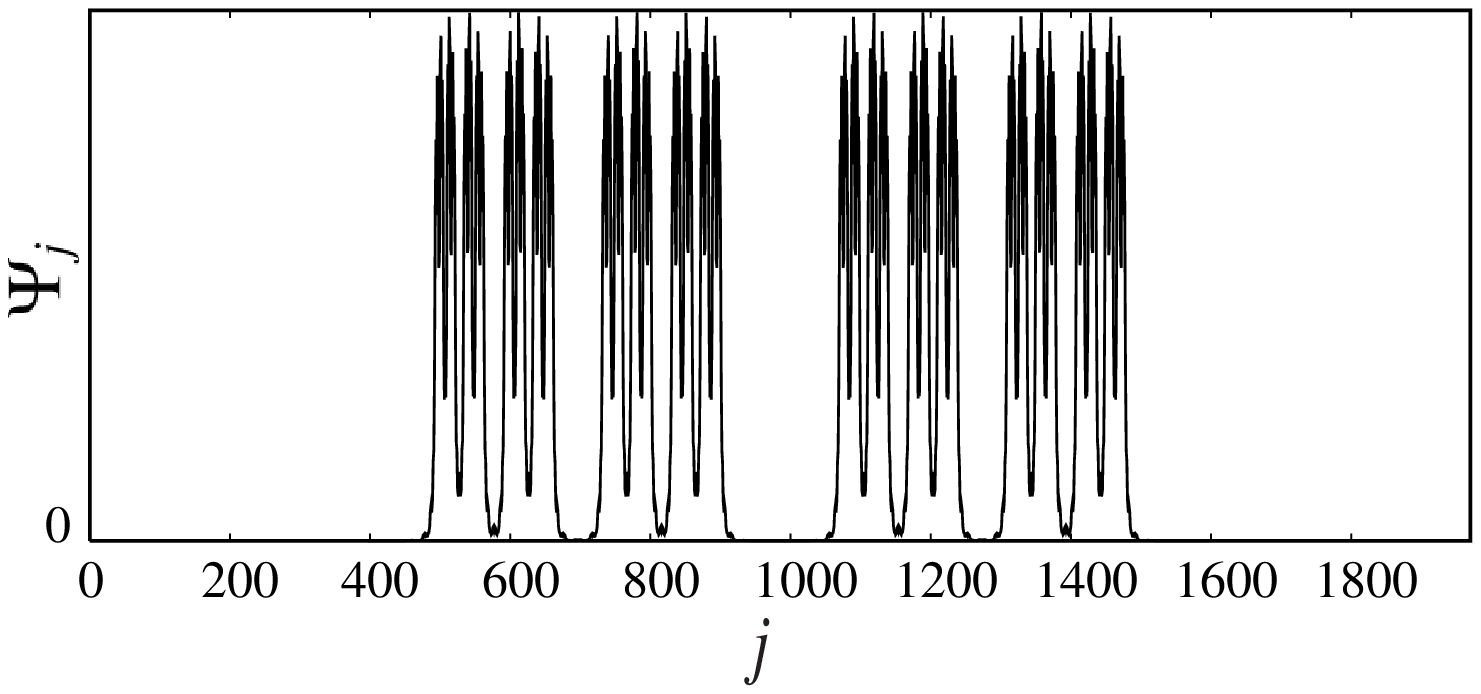,height=4cm}\end{center}
\caption{The ground state wave function of the approximant composed of
1970 atoms with $V=1.0$. $\Psi_j$ is the probability amplitude on the
$j$th site. This figure exhibits the molecular bonding state of the
eighth generation, where hierarchical internal structure is observed.}
\label{fig:wf}
\end{figure}

In the first approximation, we take only the leading effective transfer
integral $t_{\rm s}$ into account. Since the type s spacings are isolated,
the energy spectrum consists of three sub-levels at $E=0$ and
$E=\pm t_{\rm s} \; (=\mp 1)$; the central level is derived from isolated
atoms, while the two satellites from isolated diatomic ``molecules''. The
weights of the three sub-levels have the ratios
$\tau^{-1}:\tau^{-2}:\tau^{-1}$, which agree with the numerical results. To
see the trifurcation of the level derived from the molecular bonding
states, we shall consider the ``molecules'' to be the ``atoms'' of the
second generation. Due to the self-similarity of the QL, there appear three
types of effective transfer integrals between the new ``atoms'' as shown in
Fig.\ \ref{fig:dec}:
\begin{equation}
t_{\rm s}' \approx \frac{1}{2}t_{\rm m},\; t_{\rm m}' \approx
\frac{1}{2}t_{\rm l},\;{\rm and}\; t_{\rm l}' \approx
\frac{1}{2}\frac{t_{\rm l}^2}{t_{\rm s}},
\label{transfer}
\end{equation}
where the numerical factor $1/2$ appears as the square of the bonding
amplitude $1/\sqrt{2}$. Since the new parameters satisfy the same
inequalities as Eq.\ (\ref{inequality}), we can return now to the point
after Eq.\ (\ref{inequality}) and continue the discussion {\it recursively}.
Thus, we obtain a hierarchical trifurcation spectrum. The ground state is
understood to be a hierarchical composite of molecular bonding states,
as shown in Fig.\ \ref{fig:wf}.


To make a quantitative argument, we introduce two positive parameters as
the ratios between the effective transfer integrals of the $n$th
generation:
\begin{equation}
f_n = \frac{t_{\rm m}^{(n)}}{t_{\rm s}^{(n)}},\quad
g_n = \frac{t_{\rm l}^{(n)}}{t_{\rm m}^{(n)}}.
\label{recursive}\end{equation}
Eq.\ (\ref{transfer}) yields the two-dimensional map,
$f_{n+1}\approx g_n,\quad g_{n+1}\approx f_n g_n$, which converges to the
origin of the $fg$-plane. This means that the ``atoms'' are increasingly
isolated as $n$ is increased. We can linearize this 2D map with new
variables, $\ln{f_n}$ and $\ln{g_n}$, and obtain the asymptotic expression
\begin{equation}
f_n\approx g_{n-1}\sim \exp{[-c\tau_{\rm G}^n]}=\exp{[-c\tau^{n\nu}]},
\label{asymptotic}\end{equation}
with $\tau_{\rm G}=(1+\sqrt{5})/2$, i.e., the golden mean, and
$\nu=\ln{\tau_{\rm G}}/\ln{\tau}\approx 0.546$. Note that $\tau_{\rm G}$ is
the leading eigenvalue of the $2\times 2$ matrix associated with the
linearized map and $c$ a positive constant depending on $V$.

The band width $w_n$ of the subcluster of the $n$th generation is estimated
to be $w_n \approx 2|t_{\rm s}^{(n)}|$. The relation $t_{\rm s}^{(n+1)}
\approx \frac{1}{2}t_{\rm m}^{(n)}$ as in Eq.\ (\ref{transfer}) can be used
to prove $w_{n+1}/w_n \approx t_{\rm s}^{(n+1)}/t_{\rm s}^{(n)} \approx
\frac{1}{2}f_n \sim \exp{[-c\tau^{n\nu}]}$, which tends to zero as $n \to
\infty$. This is observed in Fig.\ \ref{fig:spc}. Then, it can be proved
that $w_n \sim \exp{[-c'\tau^{n\nu}]}$ with $c'$ being another positive
constant. Since the size $L$ of the ``atom'' of the $n$th generation is
proportional to $\tau^n$, we find $w_n \sim \exp{[-(L/\xi)^\nu]}$, i.e.,
a stretched exponential, with $\xi$ being a characteristic length. This is
manifestly different from the case of usual critical states exhibiting a
power-law, $w_n \sim L^{-1/\alpha}$, and has never been observed in energy
spectra of type I QLs. Note that a power-law scaling could be observed if
the ratio $w_{n+1}/w_n$ had tended to a finite value.

The above argument proves that the exponent $\alpha$ of the ground state
vanishes. This does not mean, however, that the ground state is localized,
because the ground state level is not isolated from other levels. The
length $\xi$, which decreases as $V$ is increased and diverges as $V \to
0$, represents the size of almost isolated ``atoms''. The presence of a
characteristic length implies that the ground state wave function, shown in
Fig.\ \ref{fig:wf}, is not self-similar, although it still remains critical.
We may call the ground state a {\em marginal critical state}, presence of
which is closely connected with the self-similarity of the present QL.


The exact real-space renormalization-group formalism proposed by Ashraff
and Stinchcombe \cite{AsSt88} can also be adapted to study the ground state
properties. There appears a 7D map but its asymptotic behavior is dominated
by its behavior in a 2D subspace corresponding to the $fg$-plane. The map
being reduced to the subspace is essentially the same as above. It has also
been numerically confirmed that each energy level at a band edge, where
energy levels only accumulate from one side, gives the equivalent fixed
point of the 7D map to that of the the ground state level.

The presence of marginal critical states causes vanishing of the left end
$\alpha_{min}$ of the support $[\alpha_{min}, \; \alpha_{max}]$ of the
relevant $f(\alpha)$ spectrum. This is an important feature of electronic
properties of type II QLs, and has also been confirmed for several other
type II QLs including the one whose ratio of self-similarity is
$\tau_{\rm G}^3$. The exponent $\nu$ introduced above is a proper number to
the relevant MLD class but satisfies $0 < \nu < 1$. It is surprising that
QLs derived from a single mother lattice can belong to different
universality classes of electronic properties.

The binary sequence of atoms in the atomic chain used to investigate
electronic properties of type II QLs is found to be one of the circle
sequences investigated in detail by Aubry {\it et al} \cite{AuGoLu88}. Some
of other type II QLs can also be transformed to circle sequences if
they are decorated appropriately, but a general relationship between the
type II QLs and the circle sequences awaits a further investigation. We
should remark, however, that an indication of an unusual electronic
property of a circle sequence was first reported by Luck \cite{Lu89}.

The present report is summarized as follows: i) 1D QLs can be classified
into an infinite number of MLD classes, which are distinguished from each
other by the inflation rules of their representatives, and ii) there exists
a ``type II'' MLD class whose electronic properties are characterized by
the presence of marginal critical states.

A full account of the exact real-space renormalization-group treatment of
marginal critical states will be presented elsewhere. The present results
can also be extended to 2D and 3D QLs because these QLs are related to
1D QLs through the Ammann bars or planes \cite{Lu93}, and the results will
be reported elsewhere.

\begin{table}
\begin{center}
\begin{tabular}{ccccccl}
{}& $W$ & $|$s$|$ & $|$m$|$ & $|$l$|$ & ratio & inflation rule (or
substitution rule)\\ \hline
A & $\sqrt{2}$ & 1 & {$-$} & $\tau$ & $\tau$ & s$'$ = l, l$'$ = sll \\
B & $1 + \sqrt{2}$ & 1 & $\sqrt{2}$ & {$-$} & $\tau$ & s$'$ = sm,
 m$'$ = sms \\
C & $3 - \sqrt{2}$ & 1 & $\sqrt{2}$ & $\tau$ & {$-$} & ( s$'$ = sm,
 m$'$ = sll ) \\
D & $1 + \frac{1}{2}\sqrt{2}$ & 1 & $\sqrt{2}$ & $\tau$ & $\tau$ & s$'$ = sm,
m$'$ = sl, l$'$ = sll \\
E & $\sqrt{2} + \frac{1}{2}\sqrt{2}$ & 1 & $\sqrt{2}$ & $\tau$ & {$-$} &
 ( s$'$ = sm, m$'$ = ssm, l$'$ = slsm ) \\
F & $\frac{1}{2} + \sqrt{2}$ & 1 & $\sqrt{2}$ & $\tau$ & $\tau^2$ & s$'$ =
smsl, m$'$ = smsmsl, l$'$ = smsllsmsl \\
G & $\frac{3}{2} + \sqrt{2}$ & $\tau^{-1}$ & 1 & $\sqrt{2}$ & $\tau^2$ & s$'$
= msm, m$'$ = mlmmsm, l$'$ = mlmlmmsm \\
H & $\frac{1}{2} + \frac{1}{2}\sqrt{2}$ & 1 & $\tau$ & $\sqrt{2}\tau$ &
$\tau^2$ & s$'$ = msm, m$'$ = mlmmsm, l$'$ = mlmlmmsm\\
\end{tabular}
\end{center}
\caption{Several 1D QLs associated with the silver mean,
$\tau=1+\protect\sqrt{2}$. The first three belong to the type I MLD class,
while others to type II MLD classes.}
\label{table1}
\end{table}


\begin{references}

\bibitem{St99} {\it Physical Properties of Quasicrystals}, edited by
Z. M. Stadnik, Series in Solid-State Sciences, 126 (Springer-Verlag, 1999).

\bibitem{DuKa85} For the projection method of forming a QL, see M. Duneau
and A. Katz, Phys. Rev. Lett. {\bf 54}, 2688 (1985).

\bibitem{Lu93} For 1D QLs, see J. M. Luck {\it et al.}, J. Phys. A: Math.
Gen. {\bf 26} 1951(1993) and R. L${\rm {\ddot u}}$ck, Int. J. Mod. Phys. B
{\bf 7}, 1437 (1993).

\bibitem{Ni89} For scaling of a QL, see K. Niizeki, J. Phys. A: Math.
Gen. {\bf 22}, 193 (1989).

\bibitem{KNNF} K. Niizeki and N. Fujita, cond-mat/0009422.

\bibitem{Ba91} For the MLD, see M. Baake {\it et al.}, J. Phys. A:
Math. Gen. {\bf 24}, 4637 (1991); M. Baake, math-ph/9901014.

\bibitem{KST87} M. Kohmoto, B. Sutherland and C. Tang, Phys. Rev. B
{\bf 35}, 1020 (1987); M. Holzer, Phys. Rev. B {\bf 38}, 1709 (1988).

\bibitem{NiNo86} Q. Niu and F. Nori, Phys. Rev. Lett. {\bf 57}, 2057 (1986);
Phys. Rev. B {\bf 42}, 10329 (1990).

\bibitem{AsSt88} J. A. Ashraff and R. B. Stinchcombe, Phys. Rev. B {\bf 37},
5723 (1988).

\bibitem{Ko92} M. Kohmoto, J. Stat. Phys. {\bf 66}, 791 (1992).

\bibitem{HJKP86Ko88} T. C. Halsey {\it et al.}, Phys. Rev. A {\bf 33}, 1141
(1986); M. Kohmoto, Phys. Rev. A {\bf 37}, 1345 (1988).

\bibitem{AuGoLu88} S. Aubry, C. Godr${\rm \grave{e}}$che and J. M. Luck, J.
Stat. Phys. {\bf 51}, 1033 (1988).

\bibitem{Lu89} J. M. Luck, Phys. Rev. B {\bf 39}, 5834 (1989).

\end{references}
\end{document}